\documentclass{elsart}

\usepackage{graphicx}
\usepackage{bm}
\usepackage{amsmath}

\begin{document}

\newcommand{\MgBii}{MgB$_2$}
\newcommand{\Hcii}{H_{\text{c2}}}
\newcommand{\Tc}{T_{\text{c}}}
\newcommand{\xiGL}{\xi_{\text{GL}}}
\newcommand{\xipi}{\xi_{\pi}}
\newcommand{\Dpi}{\Delta_{\pi}}
\newcommand{\Dsig}{\Delta_{\sigma}}
\newcommand{\piband}{$\pi$-band}
\newcommand{\sigband}{$\sigma$-band}
\newcommand{\Rt}{R_{\text{t}}}
\newcommand{\It}{I_{\text{t}}}

\begin{frontmatter}
\title{Scanning Tunneling Spectroscopy on\\
       Single Crystal \MgBii}

\author[unige]{M. R. Eskildsen\thanksref{cor}},
\author[unige]{M. Kugler},
\author[unige]{G. Levy},
\author[unige,karc]{S. Tanaka},
\author[ethz]{J. Jun},
\author[ethz]{S. M. Kazakov},
\author[ethz]{J. Karpinski},
\author[unige]{\O. Fischer}

\address[unige]{DPMC, University of Geneva, Switzerland}
\address[karc]{Kansai Advanced Research Center, Communications Research
               Laboratory, Japan}
\address[ethz]{Solid State Physics Laboratory, ETH Z\"{u}rich,
               Switzerland}

\thanks[cor]{Corresponding author, address:
              DPMC, Universit\'{e} de Gen\`{e}ve,
              24 quai E.-Ansermet, CH-1211 Gen\`{e}ve 4, Switzerland.
              Fax: +41 22 702 68 69.
              E-mail: morten.eskildsen@physics.unige.ch}

\begin{abstract}
We report on the results of scanning tunneling spectroscopy measurements on
single crystals of \MgBii. Tunneling was performed both parallel and
perpendicular to the crystalline $c$-axis. In the first case, a single
superconducting gap ($\Dpi = 2.2$ meV) associated with the \piband \ is
observed. Tunneling parallel to the $ab$-plane reveals an additional, larger
gap ($\Dsig \sim 7$ meV) originating in the highly two-dimensional
\sigband. Vortex imaging in the \piband \ was performed with the field
and tunnel current parallel to the $c$-axis. The vortices have a large core
size compared to estimates based on $\Hcii$, and show an absence of localized
states in the core. Furthermore, superconductivity between the vortices is
rapidly suppressed by an applied field. A comparison to specific heat
measurements is performed.
\end{abstract}

\end{frontmatter}

\section{Introduction}
Since the first report of superconductivity with a remarkably high $\Tc = 39$ K
in magnesium diboride (\MgBii) only two years ago \cite{nagamatsu}, enormous
progress has been made in the studies of this material. From a fundamental
point of view, a central issue has been to establish that \MgBii \ is a
two-gap superconductor - a concept which was introduced already in the fifties
\cite{suhl,gladstone}, and which has now found renewed relevance. Two-gap
superconductivity in \MgBii \ was first predicted theoretically \cite{liu}
and is now commonly accepted, confirmed by a large number of experiments
\cite{wang,szabo,chen,giubileo,bouquet,schmidt,junod,iavarone}.

The two gaps in \MgBii \ are associated with different parts of the Fermi
surface, which is composed of 4 seperate sheets \cite{liu,kortus}. Two of these
are coaxial cylindrical sheets parallel to $c^*$, derived from
$\sigma$-antibonding states of the boron $p_{xy}$ orbitals ($\sigma$-band). The
other two sheets are derived from $\pi$-bonding and antibonding states of the
boron $p_z$ orbitals ($\pi$-band) and are three-dimensional. Calculations of
the superconducting gap size for the two different bands yields
$\Dsig \approx 7$ meV, and $\Dpi = 1-3$ meV \cite{golubov,choi}. These values
are respectively above and below what is expected for a BCS $s$-wave
superconductor, $\Delta = 1.76 \; k_{\text{B}} \Tc = 5.9$ meV. Furthermore, and
in contrast to many materials or alloys studied earlier, the two bands in
\MgBii \ have roughly equal weight, leading to new and interesting phenomena as
we will show in the following. In this respect it has become clear, that local
spectroscopic investigations of the mixed state is an ideal way to study the
detailed nature of the superconducting state in \MgBii \ at a microscopic
level.

In this paper we summarize the present status of our scanning tunneling
spectroscopy measurements, made possible by the recent availability of high
quality \MgBii \ single crystals. The results include spectra obtained by
tunneling parallel as well as perpendicular to the crystalline $c$-axis,
showing how the tunnel direction changes the coupling to the two bands. In
addition, vortex imaging was performed with the tunnel current and magnetic
field parallel to the $c$-axis. The vortices are found to have a number of
remarkable properties: An absence of localized states, a very large vortex core
size compared to the estimate based on $\Hcii$, and a strong core overlap.
Some of these results were reported earlier \cite{eskildsen}.

\section{Experimental}
The scanning tunneling spectroscopy (STS) experiments were performed using a
home built scanning tunneling microscope (STM) installed in a $^3$He, UHV
cryostat containing a 14 T magnet \cite{kugler}. The tunneling measurements
were done using electrochemically etched iridium tips, with the bias voltage
applied to the sample, i.e. a positive bias corresponds to probing the empty
states and negative bias to the occupied states respectively above and below
the Fermi level. A tunnel resistance of $\Rt = 0.2 - 0.4$ G$\Omega$ was used,
and the differential conductance measured using a standard ac lock-in
technique.

Single crystals of \MgBii \ were grown using a high pressure method described
elsewhere \cite{karpinski}, yielding platelike samples with the surface normal
parallel to the crystalline $c$-axis. The surface area of the crystals are
roughly $0.25 \times 0.25$ mm$^2$, and the thickness varies between a few up to
$\sim 100$ microns. The critical temperature is typically $\Tc = 37 - 38.6$ K,
with a sharp transition, $\Delta \Tc = 0.4 - 0.6$ K, measured by  SQUID
magnetometry \cite{karpinski}. Measurements were done both with the tunnel
current, $\It$, parallel and perpendicular to the $c$-axis. For
$\It \parallel c$, the spectroscopy was done on the surface of an as grown
single crystal. In the case of $\It \perp c$, a relatively thick crystal was
cracked to expose a clean surface and immediately mounted and inserted into
UHV.

The upper critical field of \MgBii \ is highly anisotropic \cite{budko}, with
zero temperature extrapolations of respectively $\Hcii^{\text{c}} = 3.1$ T and
$\Hcii^{\text{ab}} = 23$ T for single crystals \cite{angst}, even though the
latter value is subject to some uncertainty. Using the Ginzburg-Landau (GL)
expression for $\Hcii = \phi_0 / (2 \pi \xi^2)$, where $\phi_0 = h/2e$ is the
flux quantum, yields an in-pane coherence length, $\xiGL^{\text{ab}} = 10$ nm.
Due to the large anisotropy,
$\xiGL^{\text{c}} = \xiGL^{\text{ab}} \times \Hcii^{\text{c}}/\Hcii^{\text{ab}}
                  \simeq 1 - 2$ nm.
An estimate of the mean free path, based on the measured residual resistivity
\cite{sologubenko} and specific heat \cite{junod}, and the calculated Fermi
velocity \cite{brinkman}, gives $l = 50 - 100$ nm, and a first estimate is thus
that the samples are in the moderately clean limit \cite{huebener}. However,
the two-band nature of \MgBii \ complicates the picture since $\Hcii$ is
dominated by the two-dimensional \sigband \ which gives rise to the large
$\Hcii$-anisotropy, while the transport properties depends on both bands. A
detailed analysis of the thermal conductivity by Solugubenko {\em et al.}
concluded that the mean free path is roughly equal for the two bands with
$l \approx 80$ nm \cite{sologubenko}.

\section{Results}
In the following we present the results of our measurements on \MgBii \ single
crystals.

\subsection{Dependence on the tunnel current direction}
The observation of a single or both superconducting gaps depends on the
orientation of the sample, since tunneling along different directions changes
the coupling to the two bands \cite{brinkman}. Roughly speaking, only bands
with a component of the Fermi velocity parallel to the tunneling direction are
observed. In particular, this means that coupling to the 2D \sigband \ is
highly suppressed for tunneling parallel to the $c$-axis, whereas the 3D
\piband \ will be probed for all tunneling directions.

\begin{figure}
\includegraphics*[width=7.5cm]{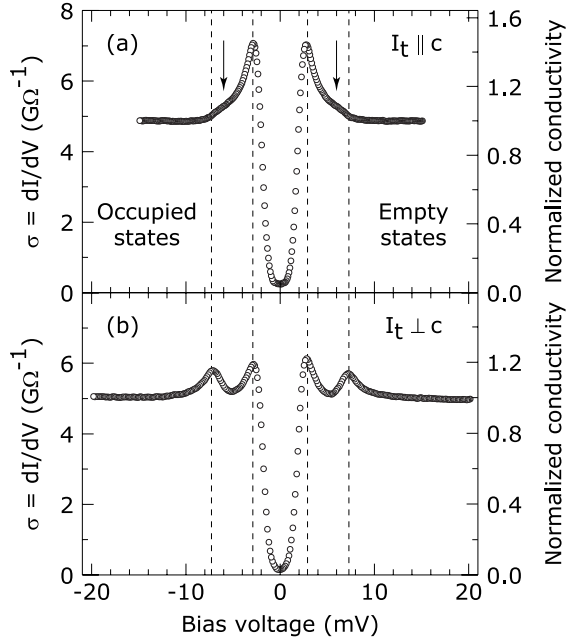}
\caption{
  Zero-field superconducting spectra of \MgBii \ obtained by tunneling
  respectively parallel (a) and perpendicular (b) to the $c$-axis. In both
  cases $T = 0.3$ K and $\Rt = 0.2$ G$\Omega$ ($U = 0.1$ V; $I = 0.5$ nA).
  \label{spectra}}
\end{figure}

In Fig. \ref{spectra}(a) we show a superconducting spectrum with
$\It \parallel c$.
This is an average of 40 spectra obtained along a 100 nm path, which shows
perfect homogeneity. One observes a single gap with coherence peaks at
$\pm 2.9$ meV associated with the \piband, and additional weak shoulders at
$\pm 6$ meV as indicated by the arrows. The shoulders are an admixture of the
gap in the \sigband.

In Fig. \ref{spectra}(b) we show an average spectrum obtained for
$\It \perp c$. For this direction of tunneling, both superconducting gaps are
present. The position of the low-energy coherence peak is unchanged as shown by
the dashed lines. The additional peak arising from the \sigband \ is located at
$\pm 7.3$ meV. This is the first report of tunneling measurements on single
crystals of \MgBii \ for both $\It \parallel c$ and $\It \perp c$, which allows
a direct correlation between the tunneling direction and the observed gap(s).
Our results agree with those of Iavarone {\em et al.}, who investigated a
number of single grains with different, but unknown absolute orientations
\cite{iavarone}.

For both tunneling directions the low zero bias conductance indicates a high
quality tunnel junction and a low noise level. True vacuum tunneling conditions
were confirmed by varying the tunnel resistance, $\Rt$, and verifying that the
spectra normalized to the conductance outside the superconducting gap collapse
onto a single curve.

\subsection{Temperature dependece for $I_{\text{t}} \parallel c$}
\begin{figure}
\includegraphics*[width=7.5cm]{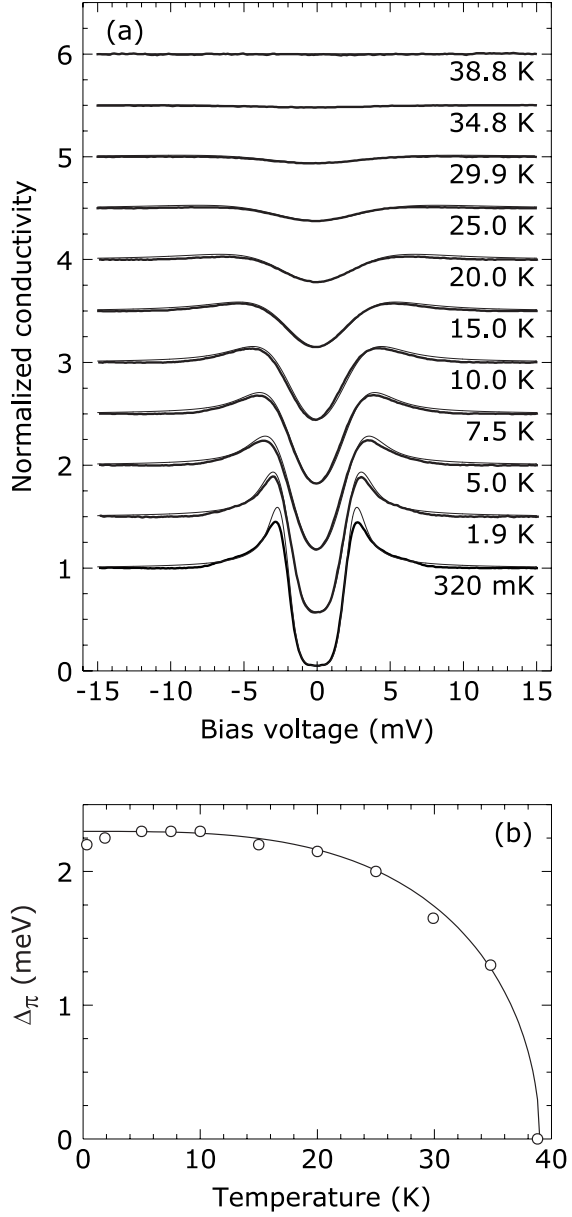}
\caption{
  (a) Spectra obtained a number of temperatures between 320 mK and $38.8$ K
  (thick lines), tunneling parallel to the $c$-axis. Each spectrum is
  normalized to the conductance at 12 meV, and offset by $0.5$ with respect to
  the previous one for clarity. The thin lines are our fits to the spectra as
  described in the text.
  (b) Temperature dependence of the superconducting gap in the \piband. The
  line is the BCS expression for $\Delta(0) = 2.3$ meV and $\Tc = 39$ K
  \cite{muhlschlegel}.
  \label{Tdep}}
\end{figure}

Tunneling parallel to the $c$-axis, we measured the superconducting spectrum
for a number of temperatures between 320 mK and $38.8$ K. As discussed above
this probes mainly the \piband. The results are presented in
Fig. \ref{Tdep}(a), and shows how the gap is gradually suppressed and finally
closes at $\Tc$.
We have fitted the measured spectra to the conductance calculated using the
BCS expression for the density of states, including a finite quasiparticle
lifetime, $\Gamma$ \cite{dynes}:
\begin{equation}
N(\epsilon,\Gamma) = N_0 \; 
  \left| \mbox{Re} \; \frac{\epsilon - i\Gamma}
                           {\sqrt{(\epsilon - i\Gamma)^2 - \Delta^2}}
  \right|.
\end{equation}
The conductance is given by
\begin{equation}
\sigma(V) = \frac{dI}{dV} \propto
  \int d\epsilon \; N(\epsilon,\Gamma) 
    \left( - \frac{\partial f(\epsilon + eV)}
                  {\partial V} \right),
\end{equation}
which we furthermore have convoluted by a Gaussian of width $\omega$ to take
into account experimental broadening. The fits are shown in Fig. \ref{Tdep}(a)
and are found to be in good agreement with the measurements, except at low
temperatures where the height of the coherence peaks is overestimated. The
extracted values of the gap are shown in Fig. \ref{Tdep}(b), yielding
$\Dpi(0) = 2.2 - 2.3$ meV. We find that $\Dpi(T)$ is in excellent agreement
with the temperature dependence of the BCS $\Delta(T)$ \cite{muhlschlegel} when
the zero temperature gap is scaled to $2.3$ meV. The lifetime parameter
$\Gamma = 0.1$ meV at low-$T$ and increases with temperature. The experimental
broadening $\omega = 0.5$ meV at base-$T$ and increases only slightly to $0.8$
meV at 10 K. For comparison the ac excitation used for the measurements was
$0.4$ meV RMS. Above 10 K the effects of $\Gamma$ and $\omega$ becomes
identical, and the latter is kept fixed.

\subsection{Vortex imaging in the \piband}
We now turn to measurements in an applied magnetic field. These measurements
were done with $\It \parallel \bm{H} \parallel c$, and hence selectively
probes the \piband. In a type-II superconductor such as \MgBii, the magnetic
field penetrates into the sample in the form of vortices each carrying one flux
quantum, and generally arranged in a periodic array: the vortex lattice. In the
core of each vortex, superconductivity is suppressed within a radius roughly
given by the coherence length, $\xi$. The vortex spacing is determined by the
applied field and the flux quantization, and in the case of a hexagonal vortex
lattice it is $d = (2/\surd 3 \; \phi_0/H)^{1/2}$.

\begin{figure}
\includegraphics*[width=7.5cm]{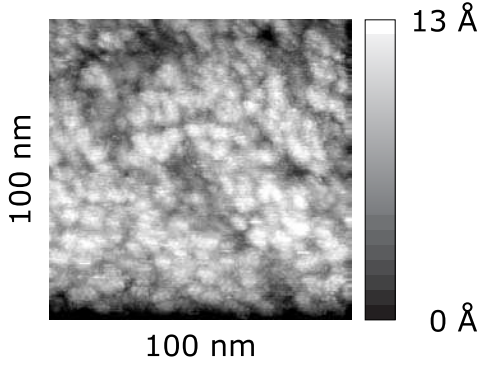}
\caption{
  STM topographic image of the as-grown surface of a \MgBii \ crystal. The RMS
  roughness is 3 \AA.
  \label{topo}}
\end{figure}

Necessary prerequisites for obtaining vortex lattice images is a high sample
homogeneity and a relatively flat sample surface. We verified that the first
requirement is fulfilled by measuring perfectly reproducible spectra in zero
field at a large number of positions along a several thousand \aa ngstr\o m
line. A topographic STM image of the sample surface is shown in
Fig. \ref{topo}.
The surface is seen to be very smooth, with an indication of a granular
structure with a characteristic size of 5 - 10 nm. No effect of this
``granularity'' is observed in the spectra.

\begin{figure}
\includegraphics*[width=15cm]{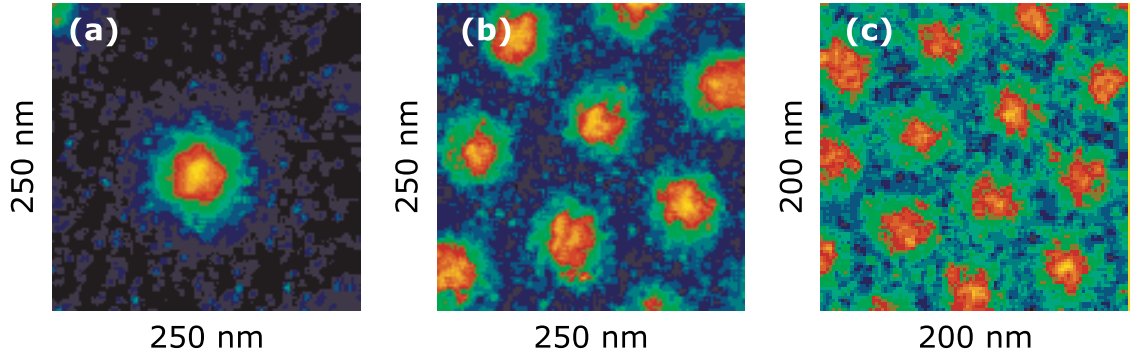}
\caption{
  (color)
  Vortices in \MgBii \ at respectively $0.05$ T(a), $0.2$ T(b) and $0.5$ T(c)
  at 2 K. The spatial variation of the conductance is shown by the color scale
  (different for each image).
  \label{vlimage}}
\end{figure}

The magnetic fields were applied at 2 K, and the system allowed to stabilize
for at least a few hours. After this time no vortex motion was observed,
indicating a fast relaxation and hence low vortex pinning in the crystal. 
In Fig. \ref{vlimage} we show STS images of vortices induced by three different
applied fields.
The images were obtained by measuring the differential conductance at zero
bias. Low values of the conductance correpond to superconducting areas, and
high values to the vortex cores. Figs. \ref{vlimage}(b) and (c) show a well
ordered hexagonal vortex lattice, with a lattice constant that corresponds to
the applied field within 10\%.

\begin{figure}
\includegraphics*[width=7.5cm]{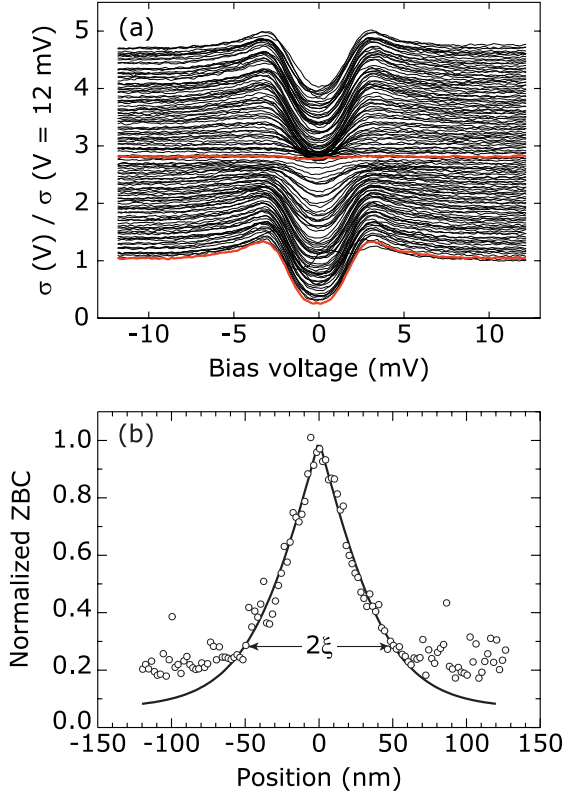}
\caption{
  (color)
  (a) 250 nm trace through the centre of the single vortex in Fig.
  \ref{vlimage}(a), with spectra recorded each 2 nm. A spectrum at the vortex
  centre together with one far from the vortex core have been highlighted in
  red for clarity.
  (b) Normalized zero bias conductance versus distance from the centre. The
  line is a fit to eq. (\ref{sigma}) in the text.
  \label{vortextrace}}
\end{figure}

The low field in Fig. \ref{vlimage}(a) is equivalent to a separation,
$d = 220 \text{ nm} \gg \xiGL$. The vortices can therefore be considered as
isolated from each other. Such isolated vortices are expected to contain
localized quasiparticle states (Andreev bound states), which should show up as
a zero bias conductance (ZBC) peak at the vortex centre \cite{hessgygi,renner},
provided that the sample is sufficiently clean to prevent these from being
smeared out by scattering. We have measured the evolution of the spectra at a
large number of positions along a trace across the vortex core. This is shown
in Fig. \ref{vortextrace}(a).
Contrary to expectations, we find that the normalized ZBC increases to one with
no indication of any localized states. Instead, the spectra at the centre of
the vortex are {\em absolutely flat}, with no excess spectral weight at or
close to zero bias. This absence of localized states is striking, considering
that $l = 5 - 10 \times \xiGL$. However, as we will show below, the coherence
length in the \piband \ is approximately 50 nm. This is much larger than the
GL estimate based on $\Hcii$, and equal to only 1 -- 2 times the mean free
path. Nonetheless, for comparison systematic studies of Nb$_{1-x}$Ta$_x$Se$_2$
with $x = 0 - 0.2$, showed that even for $\xi/l \approx 1$ some excess weight
close to zero bias was observed \cite{renner}.

The spatial extent of the vortex is shown in Fig. \ref{vortextrace}(b), where
we have plotted the normalized ZBC, $\sigma'(x,0)$ from the vortex trace. It is
immediately clear that the spatial extension of the vortex core is much larger
than the 10 nm estimated from $\Hcii$. We find that the vortex profile can be
fitted by one minus the GL expression for the superconducting order parameter:
\begin{equation}
  \sigma'(x,0) = \sigma'_0 + (1 - \sigma'_0) \times (1 - \tanh x/\xi),
  \label{sigma}
\end{equation}
where $\sigma'_0 = 0.068$ is the normalized ZBC measured in zero field. The fit
yields a coherence length, $\xi = \xipi = 49.6 \pm 0.9$ nm. Using the GL
expression to calculate the upper critical field with this value of the
coherence length, yields $H_{\text{c2}}^{\pi} = 0.13$ T. This value is in
strong contrast to the fact that we have performed vortex lattice imaging up to
$0.5$ T, and shows that a separate $\Hcii$ for the \piband \ does not exist.

\begin{figure}
\includegraphics*[width=7.5cm]{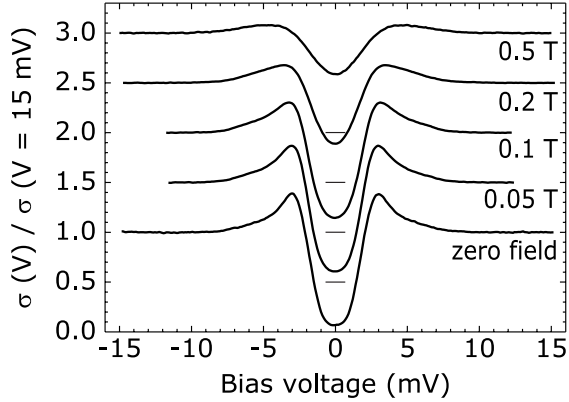}
\caption{
  Normalized spectra at 2K measured in zero field, and between the vortices for
  fields between $0.05$ T and $0.5$ T. Each subsequent spectrum is offset by
  $0.5$ with respect to the previous one, with the bars at zero bias indicating
  the respective zero conductance.
  \label{Bdep}}
\end{figure}

The large $\xipi$ has however other consequences. In Fig. \ref{Bdep} we have
plotted the normalized ZBC between the vortices.
From this it is clear that even modest fields rapidly suppress
superconductivity in the region between the vortices, seen both by an increase
of the ZBC and by a suppression of the coherence peaks outside the vortex
cores. This is consistent with earlier point contact spectroscopy measurements
\cite{szabo}, with the addition that we resolve the local behaviour on a
microscopic scale. Such a behaviour is only expected for fields close to
$\Hcii$, and corresponds to a significant core overlap.

\section{Discussion}
The vanishing of $\Dpi$ at the bulk $\Tc$ and the sustained superconductivity
despite a large vortex core overlap for $H > 0.2$ T leads to the conclusion
that superconductivity in the \piband \ is induced by the \sigband, either by
interband scattering, or Cooper pair tunneling \cite{gladstone,nakai}. This
means that by itself the \piband \ would either be non-superconducting or have
a very low $\Tc$ and upper critical field. Consequently the observed behaviour
reflects the state in the \sigband \ by an interband proximity effect, along
the lines of recent theoretical work \cite{nakai}. This is also consistent with
estimates of the coherence lengths, using the BCS expression
$\xi_0 = \hbar v_{\text{F}} / (\pi \Delta(0))$ and considering each band
separately. Taking the calculated average Fermi velocity in the $ab$-plane for
the \piband, $v_{\text{F}}^{\pi} = 5.35 \times 10^5$ m/s \cite{brinkman}, and
the measured gap value $\Dpi(0) = 2.2$ meV we get $\xi_0^{\pi} = 51$ nm, in
excellent agreement with $\xi_{\pi}$ obtained from the vortex profile. A
similar analysis for the \sigband, using
$v_{\text{F}}^{\sigma} = 4.4 \times 10^5$ m/s \cite{brinkman} and
$\Dsig(0) = 7.1$ meV \cite{iavarone} yields $\xi_0^{\sigma} = 13$ nm. This
agrees with the coherence length obtained from $\Hcii$, and reinforces the
conclusion that it is mainly the \sigband \ which is responsible for
superconductivity in \MgBii, and thus determines the macroscopic parameters
$\Tc$ and $\Hcii$. One must however bear in mind that this is a very
simple-minded analysis. A detailed theoretical treatment of two weakly coupled
superconducting bands in the presence of vortices, and the possibility of
having have different coherence lengths, have to our knowledge not yet been
presented. 

\subsection{Comparison to specific heat measurements}
\begin{figure}
\includegraphics*[width=7.5cm]{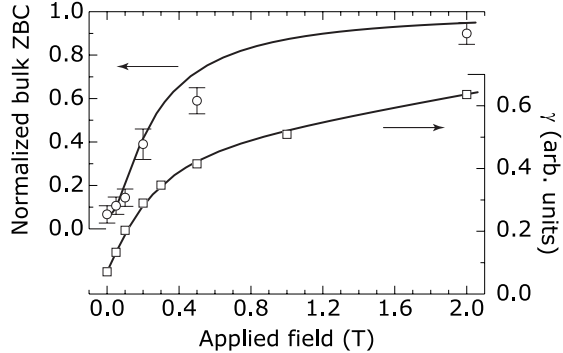}
\caption{
  Calculated ``bulk'' ZBC (left axis) and electronic specific heat, $\gamma$,
  for $\gamma^{\pi}_{\text{n}}/\gamma^{\sigma}_{\text{n}} = 0.55/0.45$ (right
  axis). The calculated values are compared to respectively the measured bulk
  ZBC (circles), and specific heat measurements on polycrystalline samples
  (squares) \protect\cite{junod}.
  \label{spheat}}
\end{figure}

The vortex core overlap explains the strongly nonlinear field dependence of the
electronic specific heat ($T$-linear term), $\gamma$ \cite{wang,junod}. In
type-II superconductors core overlap is usually negligible, and in the simplest
approximation each vortex creates the same number of quasiparticles at the
Fermi surface, hence contributing by the same amount to the specific heat. In
that case $\gamma = \gamma_{\text{n}} \times H/\Hcii$, where
$\gamma_{\text{n}}$ is the electronic specific heat in the normal state.
However, in the case of \MgBii, with strong core overlap in the \piband, the
isolated vortex assumption is violated. Constructing a very simple model for
the core overlap by
\begin{equation}
  \sigma'(\bm{r},0) =
    \sigma'_0 + 
    (1 - \sigma'_0) \times 
           \left( 1 - \Pi_i \tanh \frac{|\bm{r} - \bm{r}_i|}{\xi_{\pi}}
           \right),
  \label{overlap}
\end{equation}
where $\bm{r}_i$ are the vortex positions for a hexagonal lattice with a
density corresponding to the applied field, we can calculate the ZBC at any
position in the vortex lattice unit cell. In Fig. \ref{spheat} we compare the
calculated conductance at the midpoint between three vortices with the
measured ``bulk'' ZBC.
This shows a very good agreement, especially taking into account that there are
no free parameters in the calculation: $\xi_{\pi}$ is determined from the
vortex profile, and $\sigma'_0$ is given by the finite ZBC in the zero field
measurement. The \piband \ contribution to the specific heat can now be
calculated by averaging the normalized ZBC in one vortex lattice unit cell,
$\gamma_{\pi} = \gamma^{\pi}_{\text{n}} \; \langle \sigma'(\bm{r},0) \rangle$
\cite{nakai}. For this calculation we have set $\sigma'_0$ equal to zero. On
the other hand, the \sigband \ can be described by the usual linear field
dependence $\gamma_{\sigma} = \gamma^{\sigma}_{\text{n}} \times H/\Hcii$.
Adding the terms gives $\gamma = \gamma_{\pi} + \gamma_{\sigma}$, where
$\gamma^{\pi}_{\text{n}}/\gamma^{\sigma}_{\text{n}}$ is the relative weigth of
the two bands. The calculated field dependence of $\gamma$ is shown in Fig.
\ref{spheat}, in perfect agreement with the measured specific heat for
polycrystalline \MgBii \cite{junod}, using
$\gamma^{\pi}_{\text{n}}/\gamma^{\sigma}_{\text{n}} = 0.55/0.45$. The
distribution of weight between the two bands is in agreement with other
theoretical and experimental estimates
\cite{liu,bouquet,junod,choi,sologubenko}. Furthermore, the successful
correlation with specific heat measurements makes us confident that our results
reflect bulk properties of \MgBii.

\section{Summary}
In summary, we have presented STS data on the \piband \ in \MgBii, including
the first vortex imaging in this material. We have demonstrated the absence of
localized states in the vortex core, a very large vortex core size and a strong
core overlap. The data presents a striking experimental demonstration of the
fundamentally different microscopic properties of the two bands in \MgBii.

\section*{Acknowledgements}
We acknowledge valuable discussions and communication of data prior to
publication with F. Bouquet, M. Iavarone, A. Junod and Y. Wang, and thank
B. W. Hoogenboom and I. Maggio-Aprile for sharing their experience in STM/STS.
This work was supported by Swiss National Science Foundation.

\end{document}